\documentclass[11pt,a4paper]{article}
\pdfoutput=1
\usepackage{jheppub}

\usepackage{graphicx, caption, subcaption}  
\usepackage{dcolumn}   
\usepackage{amsmath, amssymb}       
\usepackage{color,latexsym,array,multirow,verbatim,enumerate}
\usepackage{url}
\usepackage{cancel}
\usepackage{hyperref}

\title{  Fat Jet Signature of a Heavy Neutrino at Lepton Collider}
\author[a,b]{Sabyasachi Chakraborty,} 
\author[c,d]{Manimala Mitra,}
\author[c,d]{Sujay Shil}
\affiliation[a]{Department of Physics, Florida State University, Tallahassee, FL 32306, USA,}
\affiliation[b]{Department of Theoretical Physics, Tata Institute of  Fundamental Research, Mumbai 400005, India.}
\affiliation[c]{Homi Bhabha National Institute, Training School Complex, Anushakti Nagar, Mumbai 400085, India,}
\affiliation[d]{Institute of Physics, Sachivalaya Marg, Bhubaneswar, Odisha 751005, India.}

\emailAdd{sabya@hep.fsu.edu}
\emailAdd{manimala@iopb.res.in}
\emailAdd{sujay@iopb.res.in}

\date{\today}
\abstract{We explore the discovery prospect of a  very heavy neutrino at the proposed $e^+e^-$ collider for two different c.m.energies $\sqrt{s}=1.4$ TeV and 3 TeV.  We  consider production of heavy neutrino via $s$ and $t$-channel processes, and its subsequent prompt decays leading to  semi-leptonic final states, along with significant missing energy. For our choice of masses, the  gauge boson produced from heavy neutrino decay is highly boosted, leading to a fat-jet. We carry out a detail signal and background analysis for $e^{\pm}+j_{\rm{fat}}+\cancel{E}_{T}$ final state using both cut based and multivariate techniques.  We show that a heavy neutrino of mass $600-2700$ GeV and active-sterile mixing $|V_{eN}|^2 \sim  10^{-5}$ can be probed with $5\sigma$ significance at  $e^{+}e^{-}$ collider after collecting  $\mathcal{L}=500$ $\rm{fb}^{-1}$ of data. We find the sensitivity reach at $e^{+}e^{-}$ collider is order of magnitude enhanced as compared to LHC.
 }

\keywords{Collider Physics, Seesaw Models, Heavy Neutrino Search, Beyond Standard Model Physics.}
\preprint{TIFR-TH/18-36\\
\hspace*{12.1cm} IP/BBSR/2018-14}

\begin{document}
\maketitle
\section{\label{sec:1}Introduction}
The experimental observation of neutrino oscillations in different oscillation experiments has conclusively given evidence that neutrinos have tiny eV  masses,  and non-zero mixings \cite{deSalas:2018bym}. This is a definitive indication  for the existence of   beyond the Standard Model physics (BSM physics). The solar and atmospheric mass square differences from neutrino oscillation experiments are about $\Delta m^2_{12} \sim 10^{-5}$ $\rm{eV}^2$, and $|\Delta m^2_{13}| \sim 10^{-3}$ $\rm{eV}^2$, and the mixing angles are $\theta_{12} \sim 33^{\circ}, \theta_{23} \sim 42^{\circ}$, and $\theta_{13} \sim 8^{\circ}$. Augmented with  stringent limits from Planck, the sum of light neutrino masses are bounded from above $\Sigma_i m_i \le 0.12-0.66$ eV \cite{Adam:2015rua}, where the range corresponds to different dataset considered.  A number of BSM extensions have been proposed to  explain  small neutrino masses. Few of them are the  seesaw paradigm \cite{Weinberg:1979sa, Wilczek:1979hc},  neutrino  mass generation through radiative processes \cite{Ma:1998dn, Bonnet:2012kz,Sierra:2014rxa, Zee:1980ai},  R-parity violating supersymmetry \cite{Barbier:2004ez} etc.  

Among the above, one of the most appealing framework of light neutrino mass generation is seesaw, where Majorana masses of the light neutrinos are generated from lepton number violating dimension-5 operator $LLHH/\Lambda$\cite{Weinberg:1979sa, Wilczek:1979hc}. There can be a few different variations of seesaw, Type-I \cite{Minkowski:1977sc,Mohapatra:1979ia,Yanagida:1979as,GellMann:1980vs,Schechter:1980gr,Babu:1993qv,Antusch:2001vn}, Type-II \cite{Magg:1980ut,Cheng:1980qt,Lazarides:1980nt,Mohapatra:1980yp}, and Type-III \cite{Foot:1988aq}. In Type-I and Type-III seesaw, heavy neutral leptons are included in the model. Furthermore, in Type-III,  the neutral lepton is a part of $SU(2)_L$ triplet fermionic field. In Type-II seesaw,  $SU(2)_L$ triplet  Higgs with hypercharge $Y=+2$ is included.  Both Type-I and Type-II can be embedded in  Left-Right Symmetric Model \cite{Mohapatra:1974hk, Mohapatra:1974gc, Senjanovic:1975rk} with extended gauge group. The other very popular seesaw scenario is the inverse seesaw \cite{Mohapatra:1986aw, Mohapatra:1986bd,Nandi:1985uh}, where the smallness of the light neutrino mass is protected by an enhanced lepton number symmetry of the Lagrangian.  

Most of the UV completed  seesaw models contain   Standard Model (SM) gauge singlet heavy neutrino $N$.  Depending on the  mass of the gauge singlet neutrinos and their mixings with the active neutrino states, seesaw can be tested at colliders \cite{delAguila:2008cj,Fargion:1995qb, Atre:2009rg,Cai:2017mow,Datta:1993nm,Degrande:2016aje,Mitra:2016kov,Pascoli:2018rsg,Dev:2018kpa,Deppisch:2015qwa,Bhardwaj:2018lma,Das:2017gke,Abada:2018sfh,Cottin:2018nms,Helo:2018qej,Accomando:2017qcs,Deppisch:2018eth,Kang:2015uoc,Dev:2015kca,Das:2018hph}, as well as, in other non-collider experiments, such as, neutrinoless double beta decay \cite{Mitra:2011qr,Dev:2014xea,Rodejohann:2012xd,Pas:2015eia,Gonzalez:2017mcg,Das:2017hmg,Das:2016hof}, lepton flavor violating processes $\ell_i \to \ell_j \gamma, \mu \to 3e, \mu \to e$ conversion in nuclei \cite{Abada:2007ux,Abada:2008ea}, rare-meson decays \cite{Ali:2001gsa,Mandal:2016hpr,Mandal:2017tab} etc.  Among the collider studies, LHC searches mostly focus on the charged-current production mode, i.e.,  heavy neutrino production in  $p p \to \ell^{\pm} N$, followed by  the subsequent decays of  $N$.  The smoking gun signature, that confirms the  Majorana nature of $N$  corresponds to  the same-sign di-lepton+di-jet final state \cite{Keung:1983uu,Sirunyan:2018xiv}. However the golden tri-lepton channel \cite{delAguila:2008hw} associated with missing energy is very promising, owing   to  the  smaller background rate. The active-sterile mixing $V_{lN}$ has been constrained in the range $ |V_{lN}|^2 < 10^{-5} $ for  mass of the heavy neutrino  $10\, {\rm{GeV}}<M_N<50$ GeV \cite{Sirunyan:2018mtv}. For higher masses, in particular, for TeV range $M_N$,  the LHC cross-section becomes significantly smaller. Hence, the bound on the active-sterile mixing relaxes considerably.  Other than the LHC searches, heavy neutrino can also be looked into $e^{+}e^{-}$ collider, as well as, in the $e^{-} p$ collider \cite{Mondal:2016kof,Mondal:2015zba,Lindner:2016lxq}. See 
\cite{delAguila:2005pin,delAguila:2005ssc,Das:2012ze,Banerjee:2015gca,Antusch:2017pkq,Antusch:2016ejd,Antusch:2016vyf,Antusch:2015mia,Antusch:2015gjw,Hernandez:2018cgc,Biswal:2017nfl}, for previous discussions  of the heavy neutrino searches at $e^{+}e^{-}$ collider.  Most of these works discuss the prospect of observation at $e^+ e^-$ collider for $M_N \lesssim 500$ GeV.  {For lower masses, $M_N \lesssim 500$ GeV,  ILC can probe active-sterile mixing $|V_{eN}|^2 \sim 10^{-4}$, with $\mathcal{L}=100 \, \rm{fb}^{-1}$ of data.} There is a moderate to  ultra heavy  mass range $M_N \sim $ TeV or beyond, that can further be explored in the proposed $e^{+}e^{-}$ collider Compact Linear Collider (CLIC)  \cite{Battaglia:2004mw,Linssen:2012hp, Abramowicz:2013tzc,AlipourTehrani:2254048}, {in its  higher c.m.energy run with $\sqrt{s}=1.4$ TeV, and 3 TeV.  Note, that $M_N$ upto 1 TeV can also be probed at ILC, in it's 1 TeV run.}We stress that the model signature for a very heavy $N$ is quite distinct than that of $M_N$ in the $100$ GeV mass range, that we explore in detail.  See \cite{Abramowicz:2013tzc,Contino:2013gna,Heinemeyer:2015qbu,Thamm:2015zwa,Craig:2014una,Durieux:2017rsg,Ellis:2017kfi,Dannheim:2012rn,Thomson:2015jda,Milutinovic-Dumbelovic:2015fba,Wang:2017urv,Abramowicz:2016zbo,Banerjee:2016foh} for discovery prospect of different BSM scenarios at CLIC.

In this work, we study the discovery prospect of  a heavy neutrino in the intermediate to very high mass range at  $e^{+}e^{-}$ collider. We consider two different c.m.energies $\sqrt{s}=1.4$ TeV and 3 TeV,  respectively, that are relevant for CLIC.  Contrary to the LHC, the production cross-section  of a super-heavy neutrino at $e^{+}e^{-}$ collider is fairly large. We consider two different mass ranges $M_N=600-1200$ GeV,  that can be probed at 1.4 TeV run of CLIC, and $M_N=1300-2700$ GeV, that can be discovered with  3 TeV c.m.energy.  We consider the production mode $e^{+}e^{-} \to \nu_e N$, and the subsequent decays of $N$ into an electron  $e^{\pm}$  and  $W^{\mp}$  gauge boson.  We further consider the hadronic decay modes of $W^{\pm}$. For such a heavy  $N$, the $W^{\pm}$'s are highly boosted. Hence, the   quarks from  $W^{\pm}$ are collimated, leading to a single fat-jet.   Therefore,  the final state is $e^{\pm}+j_{\rm{fat}}+\cancel{E}_T$.  We pursue an in-depth study for this final state, with both  cut-based and multivariate analysis (MVA).   We show that a heavy neutrino with mass $600-2700$ GeV and mixing $|V_{eN}|^2 \sim 10^{-5}-10^{-6}$ can be discovered with $5\sigma$ significance at $e^{+}e^{-}$ collider with $\mathcal{L} \sim 500$ $\rm{fb}^{-1}$ luminosity, which is an order of magnitude betterment as opposed to the LHC limit.

The  paper is organised as follows: in Section~\ref{int}, we discuss the interactions of  the heavy neutrino with SM particles. In Section~\ref{coll}, we discuss our model signature. Followed by this,  in Section.~\ref{cba}, we present  a detailed event analysis using  cut-based techniques for the signal and background. In Section~\ref{mva}, we optimize our search strategies using multivariate analysis (MVA), that further enhances the signal sensitivity.  The results of both the cut-based and MVA analysis are discussed in Section.~\ref{sensi}. Finally, we present our  conclusions  in Section~\ref{conclusion}. 

\begin{figure}
\centering
\includegraphics[width=0.75\textwidth]{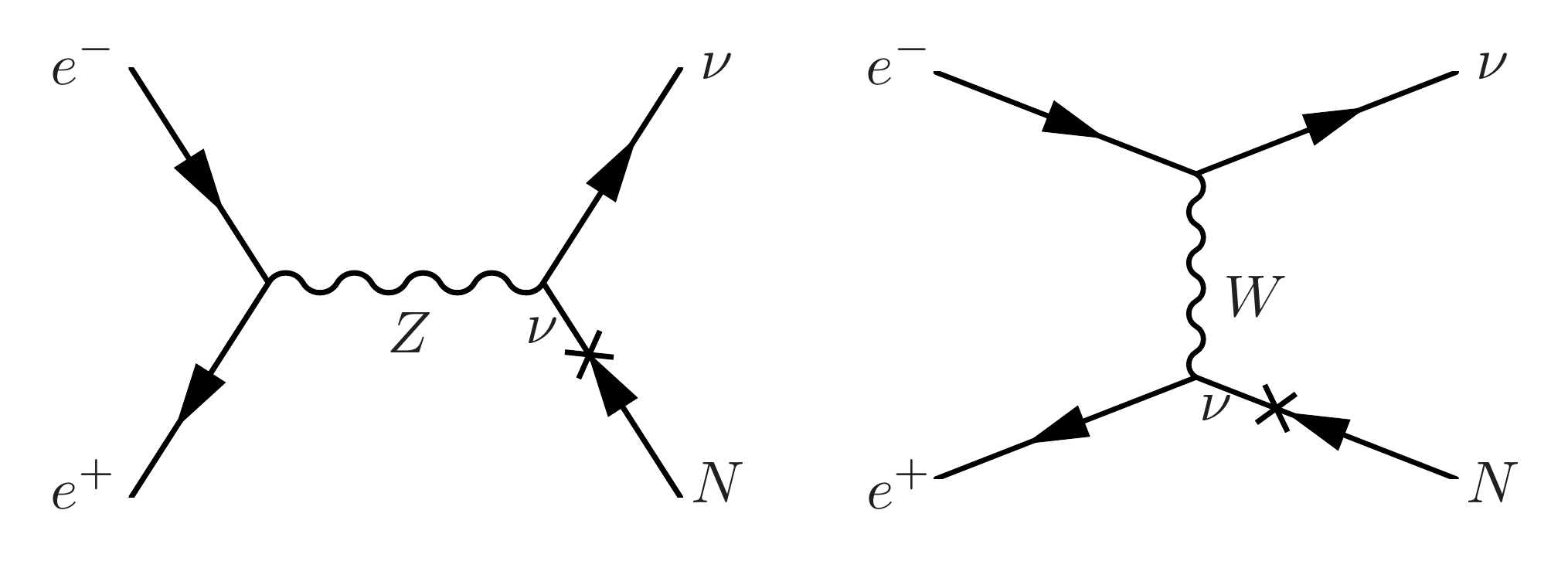}
\caption{\small{Feynman diagrams for heavy neutrino production at lepton collider. For our analysis, we consider both the $\nu_e$ and $\bar{\nu}_e$ states.}}
\label{figfeyndia}
\end{figure}
\section{Interactions   of  Heavy Neutrino {\label{int}}}
The heavy neutrino, as discussed in the introduction, can be a part of   different seesaw models,  such as, Type-I and Type-III seesaw, inverse seesaw etc. For our discussion, we follow a model independent framework, with the assumption, that the heavy neutrinos are SM gauge singlet states, and hence, do not directly interact with SM particles. Any interaction of the heavy neutrino, with the SM gauge bosons, and Higgs, is therefore governed  by its mixing with the active-neutrinos. We consider $n$-generation right-handed (RH) neutrinos $N^{\prime}_{R_{\beta}}$ (in the flavor basis), that mix with the SM light neutrinos $\nu_{L_{\alpha}}$. The light neutrinos in their flavor basis can be expressed in terms of the fields in the mass basis $(\nu_{m_i}, N_{R_j}^c$)  as follows, 
\begin{equation}
\nu_{L_{\alpha}}=U\,\nu_{L_{{i}}}+V N_{R_j}^c.
\end{equation}

In the above, $\nu_{L_i}$ refers to the active neutrinos in their mass basis, and $N_{R_j}^c$ is the conjugate-field of RH neutrino $N_R$, written in the mass basis. The matrix $U$ is the Pontecorvo-Maki-Nakagawa-Sakata (PMNS) matrix, and $V$ parametrize the mixing of the active neutrinos with the gauge singlet heavy states. Owing to the  active-sterile mixing $V$, the  heavy neutrinos $N_j$ in their mass basis  interact with the SM particles, through the charged-current, neutral-current interactions \cite{Atre:2009rg,Banerjee:2015gca}:
\begin{align}
-\mathcal{L}_{CC} \  =  \ \frac{g}{\sqrt{2}} W^-_{\mu}\bar{\ell}_i \gamma^{\mu} P_L V_{i j} {N}_j  + {\rm H.c.}\;,
\label{CC}
\end{align}
and
\begin{align}
-\mathcal{L}_{NC}  = \frac{g}{2 \cos\theta_w}  Z_{\mu} \left\{  (U^\dag V)_{ij}{\bar{\nu}}_i \gamma^{\mu} P_L {N}_j
  + {\rm H.c.} \right\}\;.
\label{NC}
\end{align}
The interaction of the heavy neutrinos with SM Higgs has the following form:
\begin{align}
-\mathcal{L}_{H}  = \frac{g M_j}{4 M_W}  H \left\{  (U^\dag V)_{ij}{\bar{\nu}}_i  P_R {N}_j
  + {\rm H.c.} \right\}\;.
\label{HIGGS}
\end{align}
In the above $M_j$ represents the mass of the heavy neutrino $N_j$. We consider a diagonal basis for the charged leptons, and hence no further mixing from charged lepton sector enters in  Eq.~(\ref{CC}). 
The partial decay widths of different decay modes have the following expression: 
\begin{align}
\Gamma(N\to \ell^- W^+) & \ = \ \frac{g^2}{64\pi}|V_{\ell N}|^2\frac{M_N^3}{M_W^2}\left(1-\frac{M_W^2}{M_N^2}\right)^2\left(1+2\frac{M_W^2}{M_N^2}\right), \label{partial1}\\
\Gamma(N\to \nu_\ell Z) & \ = \ \frac{g^2}{128\pi}|V_{\ell N}|^2\frac{M_N^3}{M_W^2}\left(1-\frac{M_Z^2}{M_N^2}\right)^2
\left(1+2\frac{M_Z^2}{M_N^2}\right), \label{partial2} \\
\Gamma(N\to \nu_\ell H) & \ = \ \frac{g^2}{128\pi}|V_{\ell N}|^2\frac{M_N^3}{M_W^2}\left(1-\frac{M_H^2}{M_N^2}\right)^2. \label{partial3}
\end{align}
For the heavy neutrino  significantly massive  than SM gauge bosons and Higgs, i.e., $M_N\gg M_W, M_Z, M_H$,  the  branching ratio is approximated as $\text{Br}(N\to \ell^{\pm}W^{\mp})$:$\text{Br}(N\to \nu_{\ell}Z)$:$\text{Br}(N\to \nu_{\ell}H)$ = $2:1:1$. We show the variation of branching ratio with mass of $N$ in Fig.~\ref{fig:crossbr}.  For $M_N \ge 600$ GeV, which is of our interest, the leading branching ${\rm{Br}}{{(N \to \ell W)} }\sim 50\%$. This has significant impact in our choice of final states, as will be cleared from the next section. 
\subsection{Production and Decay at a Lepton Collider \label{coll}}
\begin{figure}[h]
\centering
\includegraphics[width=\textwidth]{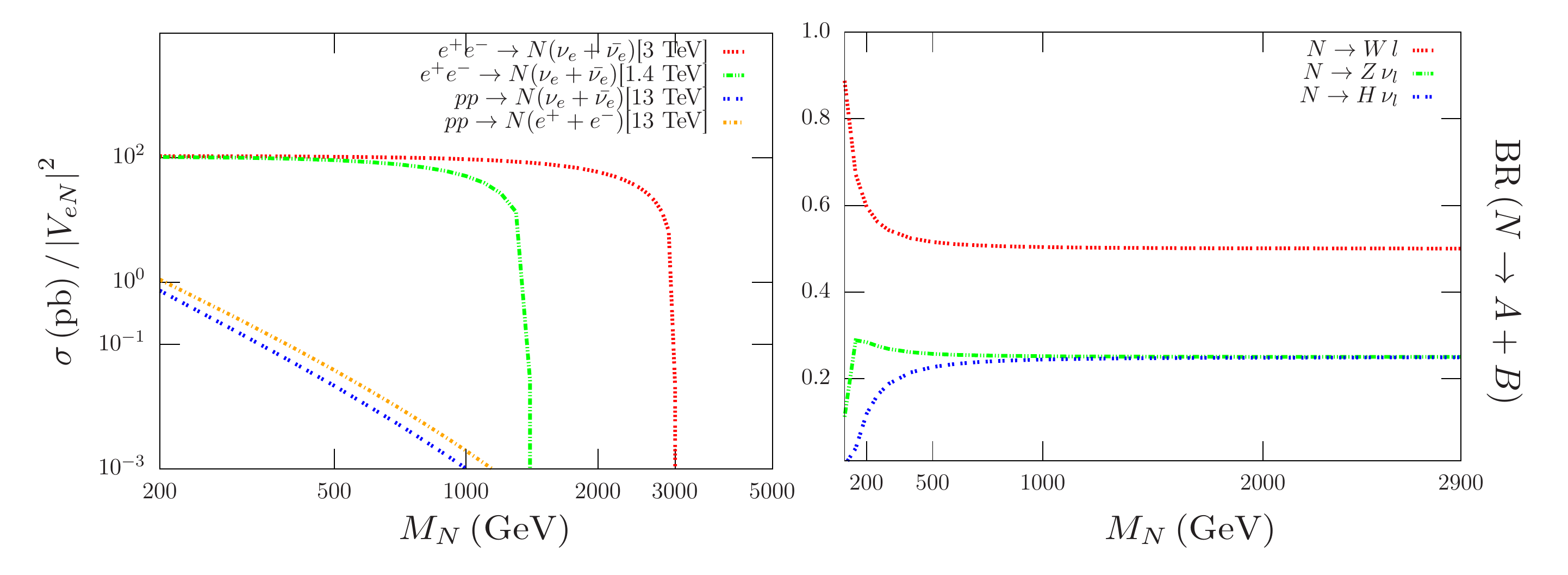}
\caption{\small{Left panel: We plot the production cross section of the heavy neutrino normalized by the active-sterile mixing parameter. We also compare our results for the compact linear collider with the LHC. Right panel: Variation of branching ratio of  $N$ vs mass. The different decay modes are  $N \to \nu_e Z, N \to  l^{\pm}W^{\mp}$ and $N \to \nu_e H$ states. }}
\label{fig:crossbr}
\end{figure}

The heavy neutrino interacts with the charged leptons, and SM gauge bosons. Due to  the interaction of the heavy neutrinos $N_j$ with $l^{\pm}-W^{\mp}$, and $\nu_{\ell}-Z$,  $N_j$ can be produced at a lepton collider.  The Feynman diagram for the production process $e^{+}e^{-} \to \nu_e N$ is shown in Fig.~\ref{figfeyndia}, and the cross-section is given in the left panel of Fig.~\ref{fig:crossbr}, for c.m.energies $\sqrt{s}=1.4$ and $3$ TeV. For comparison, we also show the production cross-section at LHC, with 13 TeV c.m.energy for  both  the channels $p p \to  e^{\pm}  N$ and $ p p \to \nu_e N$.   For hundred GeV-TeV mass range $200\, {\rm{GeV}} < M_{N} < 2900$ GeV, the normalised cross-section at a lepton collider varies from $\sigma \sim (10^2-6.7$) pb, which is larger than the production cross-section at LHC by at least $\mathcal{O}({10^2})$. To probe heavier $M_N$ at LHC, relatively large partonic c.m.energy is required. The fall in the cross-section for higher $M_N$ occurs due to the drop of the pdf. Furthermore, the channel $p p \to \nu_{\ell} N$ suffers additional suppression as compared to $e^+e^- \to \nu_{\ell} N$, due to smaller electromagnetic coupling.

The   channel $e^{+} e^{-} \to \nu_e N$ has also  been explored before in \cite{Banerjee:2015gca} for lower c.m.energies $\sqrt{s} = 250, $ and $500$ GeV. It has been inferred that a mixing down to $|V_{eN}|^2  \sim 10^{-4}$ can be  probed at a linear collider upto  $M_{N}=400$ GeV with 100 $\rm{fb}^{-1}$  of data. Recently, 13 TeV LHC searches looked for the conventional di-lepton+di-jet signature \cite{Sirunyan:2018xiv}, but also for  the golden channel tri-lepton  associated with missing energy  $p p \to \ell N \to \ell^{\pm}\ell^{\mp}\ell^{\pm}+\cancel{E}_T$\cite{Sirunyan:2018mtv}. While for relatively lower mass  $10 \, {\rm{GeV}}< M_{N} < 50$ GeV,  the bound on the active-sterile mixing is $|V_{e N}|^2 \lesssim 10^{-5}$\cite{Sirunyan:2018mtv},  and for $M_N\sim 100$ GeV, this is about $10^{-2}$,  for medium mass  range $M_{N} \gtrsim 500$GeV,   the constraint is significantly relaxed.  Almost no constraint from collider searches appears for  $M_N $ in the  TeV range. The cross-section at a lepton collider, on the other hand  is large even for a  heavier  neutrino mass, that is within the kinematic threshold. Hence,  the heavy neutrino of mass several hundred GeV or TeV  should have higher discovery prospect at a linear collider.  For the  analysis that we pursue in this work, we focus on the moderate to high mass regime of the heavy neutrino, starting from 600 GeV, upto around 3 TeV. 

Subsequent decay of the heavy neutrino produces a number of final states, that can be probed in the lepton collider. 
\begin{itemize}
\item
$e^{+} e^{-} \to \nu_\ell N \to \nu_{\ell} \ell^{\pm} W^{\mp} \to \ell j j + \cancel{E_T} $\;,
\item
 $e^{+} e^{-} \to \nu_\ell N \to \nu \nu  Z \to  j j + \cancel{E_T}\;,\  \ell^{+}\ell^{-}+\cancel{E_T} $\;,
\item
$e^{+}e^{-} \to \nu_\ell N \to \nu \nu H \to b \bar{b}+ \cancel{E_T}\;, \ \tau^{+} \tau^{-} + \cancel{E_T}$\;,
\end{itemize}
For very high mass regime of the heavy neutrino, the produced gauge bosons will be boosted. Hence, the jets from the gauge boson decay would  be collimated, leading to fat-jet.  We consider the channel with the highest branching ratio of $N$, i.e.,  $N \to \ell^{\pm} W^{\mp}$ ( with $\ell=e^+, e^-)$, and hadronic decays of the $W^{\pm}$.  Therefore, our model signature is 
\begin{center} 
\begin{itemize}
\item
$e^{+} e^{-} \to \nu_e N \to e^{\pm} W^{\mp} \nu_e \to e^{\pm} + \cancel{E_T}+ j_{\rm{fat}}$
\end{itemize}
\end{center}

In our   analysis, we include both the production modes $e^+e^- \to \nu_e N$, and $e^+e^- \to \bar{\nu}_e N$.  For simplicity, in the above we consider only one decay channel of the heavy neutrino $N \to e^{\pm} W^{\mp}$.  This occurs if the active-sterile mixing $V \simeq I$ nearly diagonal. However, for non-diagonal mixing matrix, $N$ can decay to all the three flavors  $e,\mu,\tau$. The $\tau$ will again decay either hadronically or leptonically. Therefore, in the more generic scenario with all the flavors, the final state leptons would be  $e^{\pm}$, and $\mu^{\pm}$.

\section{Collider Analysis \label{analysis}}
We perform both the cut based and multivariate analysis to probe heavy  neutrinos at  collider.  To simulate the signal events, we write the interactions of the heavy neutrinos (Eq.~(\ref{CC})--Eq.~(\ref{HIGGS}))  in FeynRules \cite{Christensen:2008py,Alloul:2013bka}. The generated  Universal FeynRules Output (UFO) \cite{Degrande:2011ua} model files are then fed into  Monte-Carlo (MC) event generator MadGraph5 aMC@NLO \cite{Alwall:2014hca} to generate   event sample for the analysis. The partonic events are then passed through Pythia8 \cite{Sjostrand:2001yu} for showering and hadronization, and detector simulation has been carried out  with Delphes-3.4.1 \cite{deFavereau:2013fsa}, with the ILD card.  We use  Cambridge-Achen jet clustering algorithm \cite{Dokshitzer:1997in} to form jets, where we  consider the radius parameter $R=1.0$.  For the signal, we consider the active-sterile mixing $|V_{eN}|=0.01$, so that heavy neutrino $N$ has large decay width ($ \Gamma_N=  2.77\times 10^{-2} \, \rm{GeV}-2.58 $ GeV  for $M_N=600-2700$ GeV ), and  the decay of $N$ occurs within the detector. We generate background as $e^{\pm}+ \nu_e/\bar{\nu}_e + j j$  in MadGraph5 aMC@NLO, and follow the same set of tools for analysis. The background $e^{\pm} \nu_e/\bar{\nu}_e j j$ arises from $W^{\pm} W^{\mp}$, but also from other production process ($t$ channel mediated diagrams, off-shell gauge boson contributions etc).  In our analysis, we omit the $\tau \nu_{\tau} j j$ background, as after taking into account the leptonic branching ratios, the cross-section becomes order of magnitude smaller  ($\sigma \sim 12\, \rm fb$). Moreover, the electron, that originates from $\tau$ decay largely fail to pass our selection criterion.

We split the analysis in two different categories, a)  heavy neutrino with mass 600-1200 GeV can be probed with $\sqrt{s}=1.4$ TeV c.m.energy, b) more massive heavy neutrino upto mass $M_N \sim 3.0$ TeV can be probed with $\sqrt{s} = 3$ TeV.  We reiterate that  the final state that we demand has a  single isolated charged lepton $e^{\pm}$, one fat-jet $j_{\rm{fat}}$ with jet radious $R=1.0$, and missing transverse energy $\cancel{E_T}$.

\subsection{Cut based Analysis: \label{cba}}
\subsubsection{$M_{N}=600-1200$ GeV with $\sqrt{s}=1.4$\, TeV: \label{medrange}}
\begin{figure}[h!]
\centering
\includegraphics[width=\textwidth]{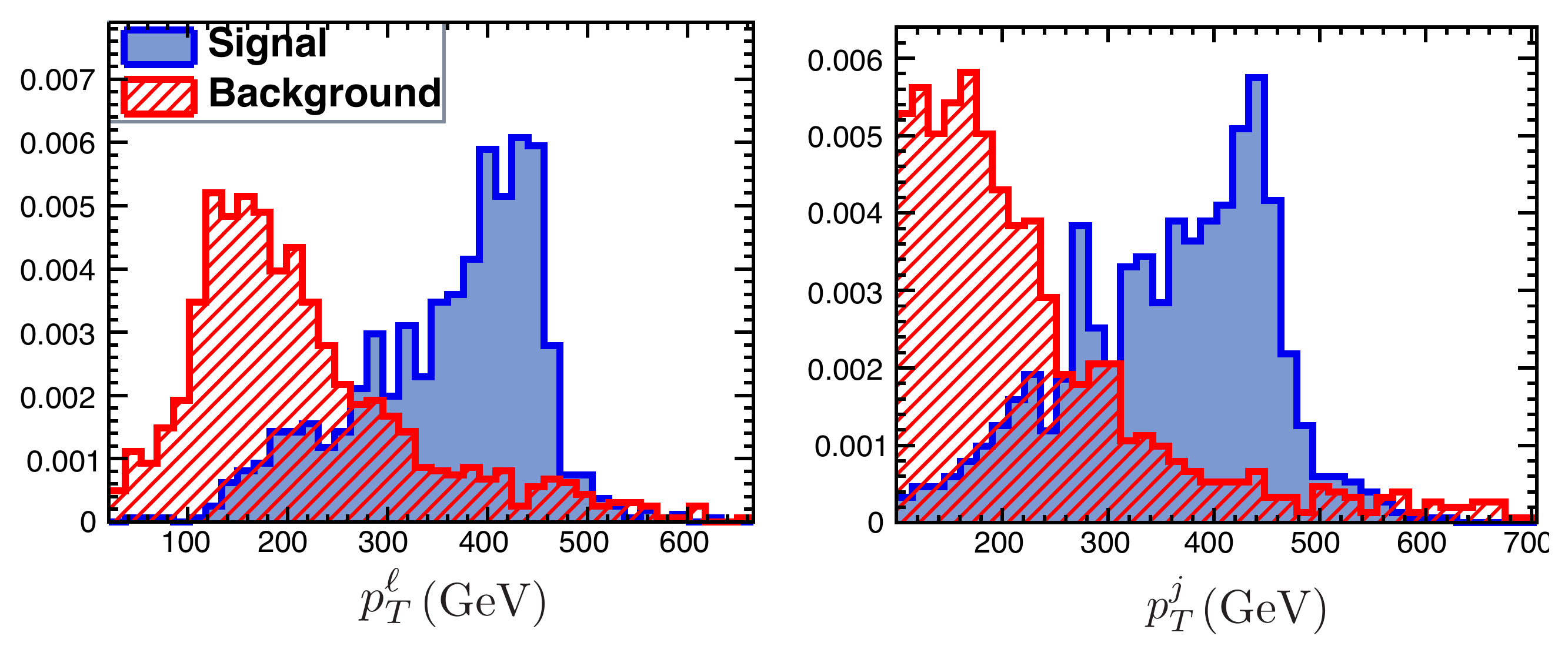}
\caption{\small{The $p_T$ distribution of lepton and fat-jet for the heavy neutrino mass $M_N=900$ GeV.}}
\label{fig:distri1}
\end{figure}

\begin{figure}[h!]
\centering
\includegraphics[width=\textwidth]{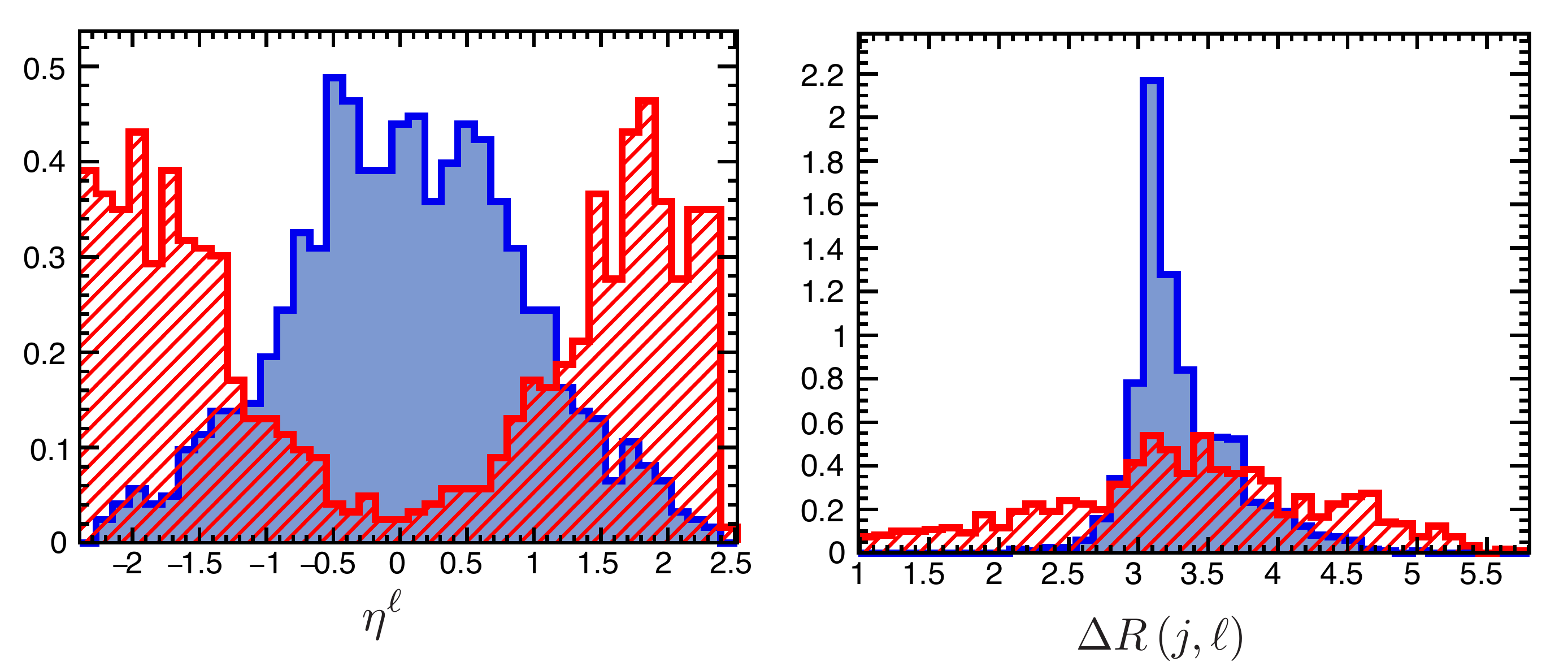}
\caption{\small{ The pseudo-rapidity $\eta^{\ell}$  distribution of charge lepton (left panel) and $\Delta R$ separation between jet and lepton  (right panel)   for  heavy neutrino mass $M_N=900$ GeV. The peak in $\Delta R(j,\ell)$ for background sample arises primarily due to $W^{\pm} W^{\mp}$ contribution. } }
\label{fig:distri2}
\end{figure}

\begin{figure}[h]
\centering
\includegraphics[width=\textwidth]{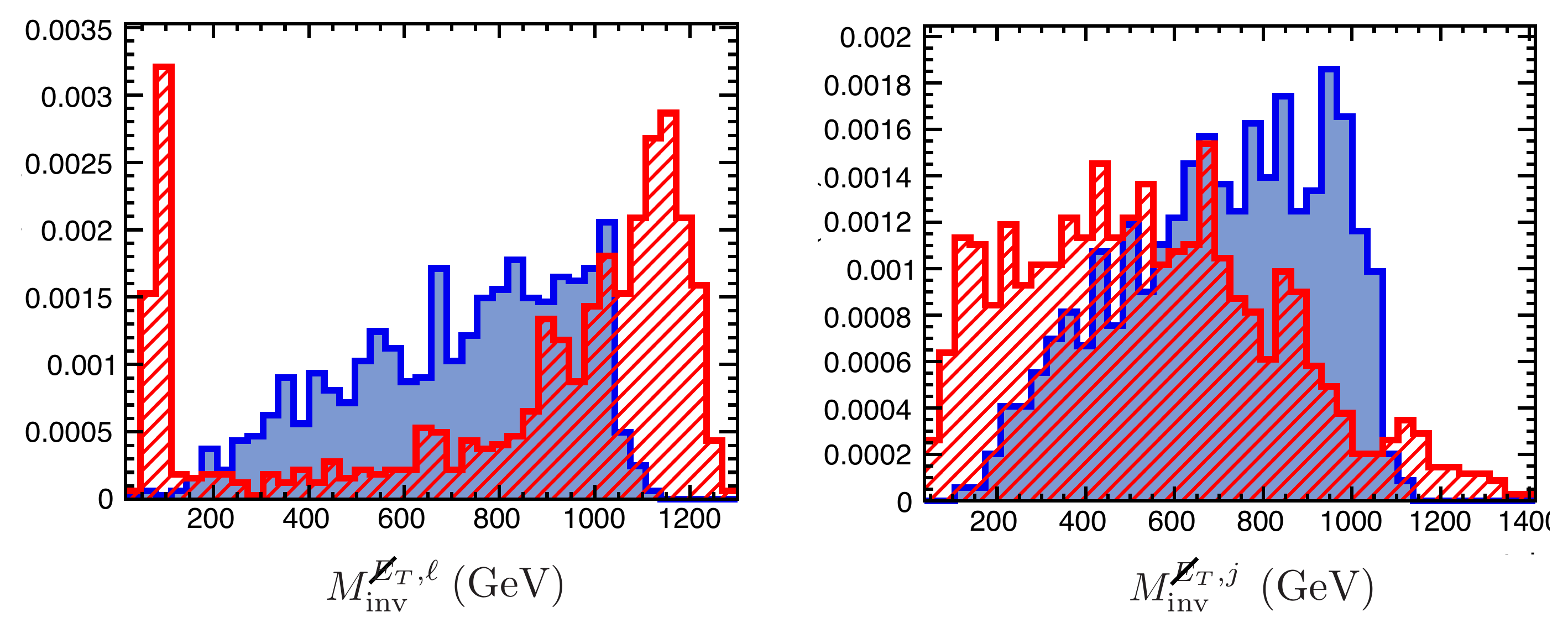}
\caption{\small{ The invariant mass distribution of the lepton, and MET (left panel) and for jet, MET (right panel)   for  heavy neutrino mass $M_N=900$ GeV.}}
\label{fig:distrib3}
\end{figure}

\begin{figure}[h]
\centering
\includegraphics[width=0.55\textwidth]{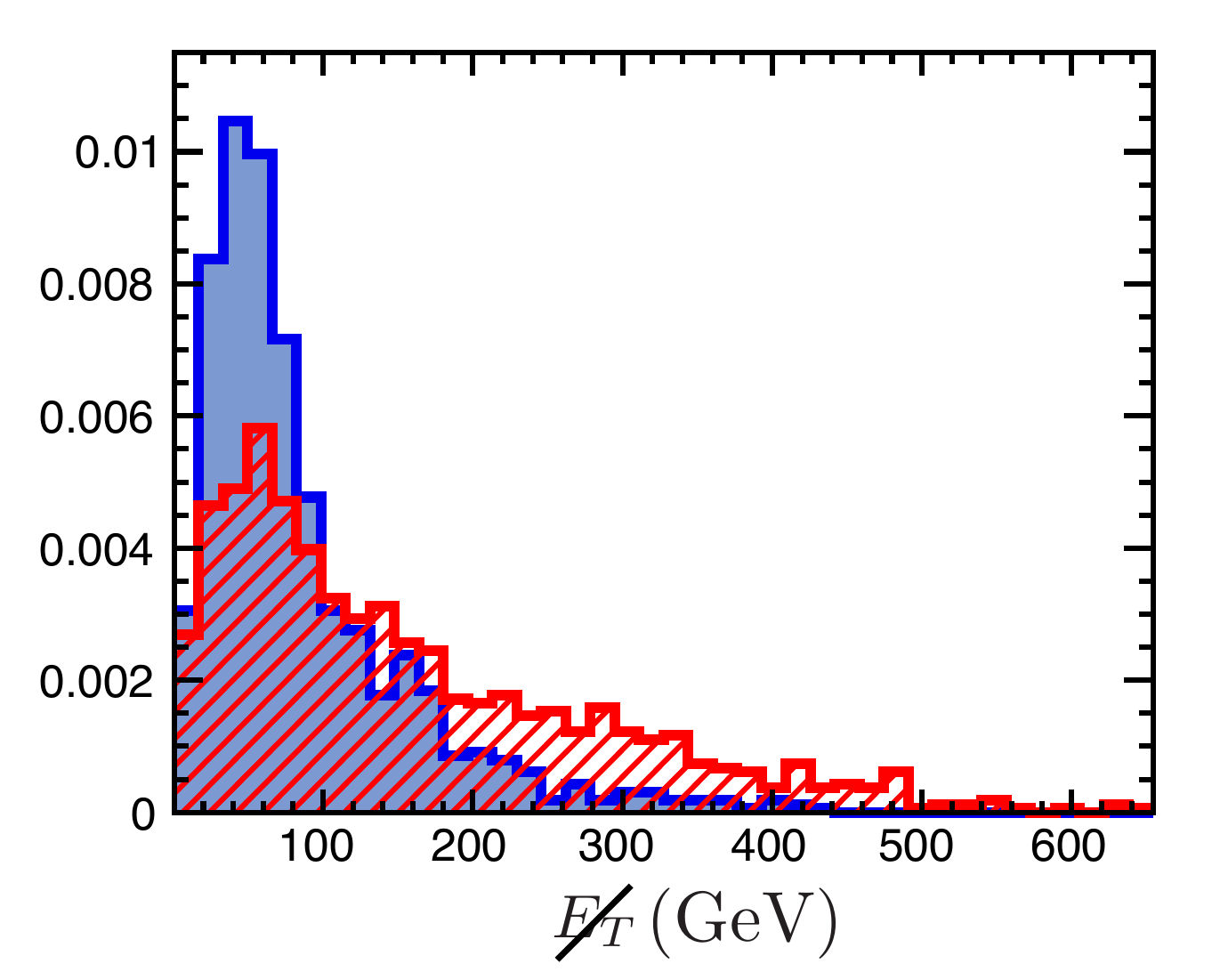}
\caption{\small{ The missing transverse energy $\cancel{E_T}$ distribution for heavy neutrino mass $M_N=900$ GeV.}}
\label{fig:distrib4}
\end{figure}

At the c.m.energy $\sqrt{s}=1.4$ TeV, heavy neutrino mass upto ${M}_N\sim 1400$ GeV can be explored kinematically.  As can be seen from Fig.~\ref{fig:crossbr}, the fall in the cross-section occurs near the kinematic threshold.  However, a wide range of masses starting from few hundred GeV upto TeV have fairly large production cross-section. As an illustrative example, we consider $M_N=900$ GeV. For this choice of mass, the production cross-section is  $\sigma(e^+e^- \to \nu_e N)=17.8$ pb, for the active-sterile mixing $|V_{eN}|=I$.  Production cross-section being proportional to $|V_{eN}|^2$, falls down to $ \sigma(e^+e^- \to \nu_e N)=1.78 $ fb for mixing $|V_{eN}|=10^{-2}$. In the subsequent analysis, we consider the above mentioned value of the active-sterile mixing, which  is in agreement with the experimental bounds from  LHC,  in the mass region that we consider. The lepton and fat-jet in the signal and background have different features in their kinematic distributions, that we utilise  for background reduction.  The distribution of various kinematic variables has been shown  in Fig.~\ref{fig:distri1}, Fig.~\ref{fig:distri2}, Fig.~\ref{fig:distrib3}, and Fig.~\ref{fig:distrib4}, both for   the signal (for sample mass point  $M_N=900$ GeV) and SM background. 
As can be seen from Fig.~\ref{fig:distri1}, the resulting lepton and the fat-jet  that originate from the decay of  heavy neutrino,  have fairly large transverse momentum, with the peak occurring around  $p_T \sim 400$ GeV.  On the other hand, the lepton and fat-jet  from background  have relatively lower $p_T$,  as it is not originating from a very heavy  state as signal. Therefore, the choice of high-$p_T$  for the lepton and also for the fat-jet removes a large fraction of the backgrounds. We  divide our analysis into two separate  segments,  one for $M_N=600-900$ GeV,  and   another for $M_N=1000-1200$ GeV. The produced  lepton and fat-jet, therefore, have relatively larger  $p_{T}$.  This motivates us to use a relatively strong cut on  charged lepton $p_T$ for   $M_N=1000-1200$ GeV,  as compared to $M_N=600-900$ GeV, and achieve better signal sensitivity.

In addition to the $p_T$ of lepton and jet, we also use a strong cut on the pseudo-rapidity $\eta^l$ of the lepton. The distribution of $\eta^\ell$ for signal and background, as can be seen from the left panel of  Fig.~\ref{fig:distri2} shows sharp contrast. For the signal, the lepton is produced in the central region, while  for background, the  peak occurs at $\eta^{\ell}$ far from zero.  In the  $e^{\pm}\nu_e/\bar{\nu}_e jj$ background the $W^{+}W^{-}$ pair production contribution is large ($\sigma \sim 73\, \rm fb$) as  compared to the other contributions.  For higher c.m.energy,  $W^{+}W^{-}$ pair produce more frequently along the beam line. This results in the non-central feature of the lepton from the background.  

In Fig.~\ref{fig:distri2} (right panel), we show the $\Delta R$ separation between the charged lepton and the fat-jet.  For $\sqrt{s}=1.4$  TeV,  the heavy neutrino of mass $M_N=900 $  GeV does not have very large momentum as compared to the case when $M_{N}$ has smaller value. Therefore,  the decay products of  $N$ will have large separation and peak of   $\Delta R$ occurs around $\Delta R(j,\ell)\sim 3.0$. For smaller value of $M_{N}$, heavy neutrino associates with larger momentum. Hence the separation would be smaller, and the peak of $\Delta R(j,\ell)$ will shift towards smaller values. For the background,   the separation between lepton and fat-jet  arising from $W^{+}W^{-}$ sample is large. However,  for other background contributions, this feature does not hold.  Therefore, for the background,  the peak of  $\Delta R$ distribution around  $\Delta R(j,\ell)\,=\,3.0$ is smaller, and primarily arises due to $W^{+}W^{-}$ pair production. We implement  a large separation cut between jet and charge lepton to remove the background. For our mass choice, the lepton and fat-jet are well separated, having large $\Delta R(j,\ell)$.
 
 For completeness, we also show the distribution of invariant mass between  MET and  lepton (jet). The invariant mass between two particles is large when their  angular separation is large. Relatively lighter heavy neutrino state will have large momentum. In this case, the produced $W$s will be aligned along the direction of $N$.  Therefore, for  lower $M_{N}$, the  angular separation between MET and jet, originated from $W$ decay is large, that results in a larger  invariant mass  $M(\cancel{E}_T,j)$. As a result,  we implement a higher  cut on $M(\cancel{E}_T,j)$ for relatively lower $M_{N} \sim 600-900$ GeV as compared to the higher mass range 1000-1200 GeV. $M(\cancel{E}_{T},\ell)$ also have similar  feature.  However,  we implement same cut for the entire mass range. For the background distribution, invariant mass $M(\cancel{E}_{T},\ell)$ have another peak near $80$ GeV, that occurs primarily  due to $W^{+}W^{-}$ contribution.
The source of missing energy is different in signal and background topologies. For the signal, $\cancel{E}_T$ is large for  relatively lower $M_N$. We show the distribution in Fig.~\ref{fig:distrib4}. We demand $\cancel{E}_T < 150.0$ throughout our analysis. Below we list  different cuts that we implement.  We have mildly  optimised our cuts for the  two different mass regions $M_{N}=600-900$ GeV (referred as CBA-{{I}}), and  $1000-1200$ GeV  (referred as CBA-{{II}}) for cut-based analysis. The cuts are constructed in such a way that we achieve the best signal significance. 
\vskip 0.5cm
\underline{\bf{ CBA-I for $M_{N}=600-900$ GeV}}
\begin{itemize}
\item $\bf{C1:}$ Transverse momentum for $e^{\pm}$: $p_{T} > 200$ GeV.
\item $\bf{C2:}$ Transverse momentum of the fat-jet:  $p_{T} > 200$ GeV.
\item $\bf{C3:}$ Transverse missing energy: $\cancel{E_T} < 150.0$ GeV.
\item $\bf{C4:}$ Pseudo-rapidity of $e^{\pm}$: $-1.0 < {\eta^\ell} < 1.0$.
\item $\bf{C5:}$ Jet-lepton separation: $ 2.8 < \Delta R(j,\ell) < 3.8. $.
\item $\bf{C6:}$ Invariant mass of transverse missing energy and lepton:  $150\, {\rm{GeV}} <M(\cancel{E}_T,\ell)  < 950$ GeV.
\item $\bf{C7:}$ Invariant mass of transverse missing energy and jet:  $M(\cancel{E}_T,j)> 600.0$ GeV.
\end{itemize}
We again optimize the cuts in the different mass window as:\\

\underline{\textbf{CBA-II for $M_{N}=1000-1200$ GeV}}
\begin{itemize}
\item $\bf{C1:}$ Transverse momentum for $e^{\pm}$: $p_{T} > 350$ GeV.
\item $\bf{C2:}$ Transverse momentum of the fat-jet:  $p_{T} > 350$ GeV.
\item $\bf{C3:}$ Transverse missing energy: $\cancel{E}_T < 150.0$.
\item $\bf{C4:}$ Pseudo-rapidity of $e^{\pm}$: $-1.0 < {\eta^\ell} < 1.0$.
\item $\bf{C5:}$ Jet-lepton separation: $ 2.8 < \Delta R(j,\ell) < 3.8 $.
\item $\bf{C6:}$ Invariant mass of transverse missing energy and lepton:  $150\, {\rm{GeV}} <M(\cancel{E}_T,\ell)  < 950$ GeV.
\item $\bf{C7:}$ Invariant mass of transverse missing energy and jet:  $M(\cancel{E}_T,j) > 400.0$ GeV.
\end{itemize}

Below, we discuss in   detail  heavy neutrino searches for  $\sqrt{s}=3$ TeV. 
\subsubsection{$M_{N}=1300-2700$ GeV with   $\sqrt{s}=3$\, TeV:}

\begin{figure}[h]
\centering
\includegraphics[width=\textwidth]{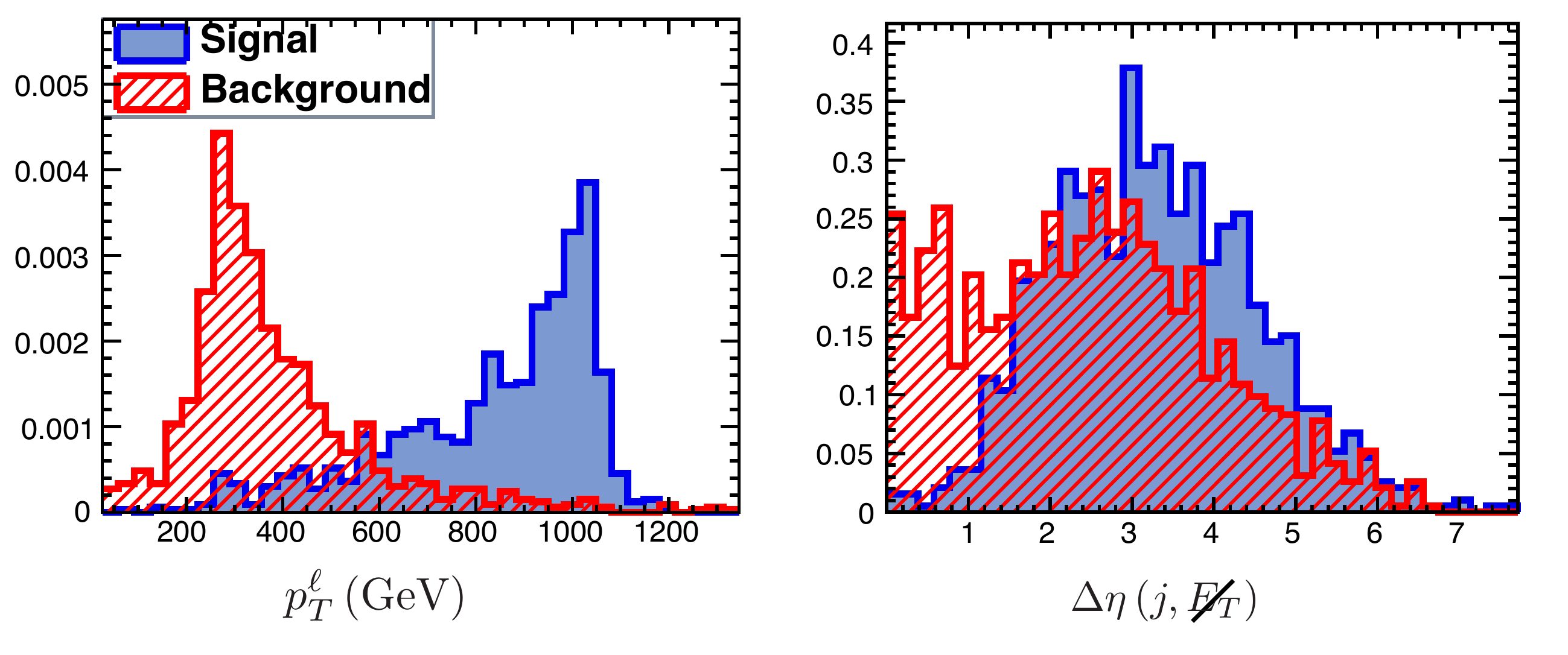}
\caption{\small{ The $p_{T}$ distribution of the charged lepton (left panel) and pseudo-rapidity separation between jet and $\cancel{E_{T}}$ (right panel)  for  heavy neutrino mass $M_N=2100$ GeV.}}
\label{2100_fig1}
\end{figure}

\begin{figure}[h]
\centering
\includegraphics[width=\textwidth]{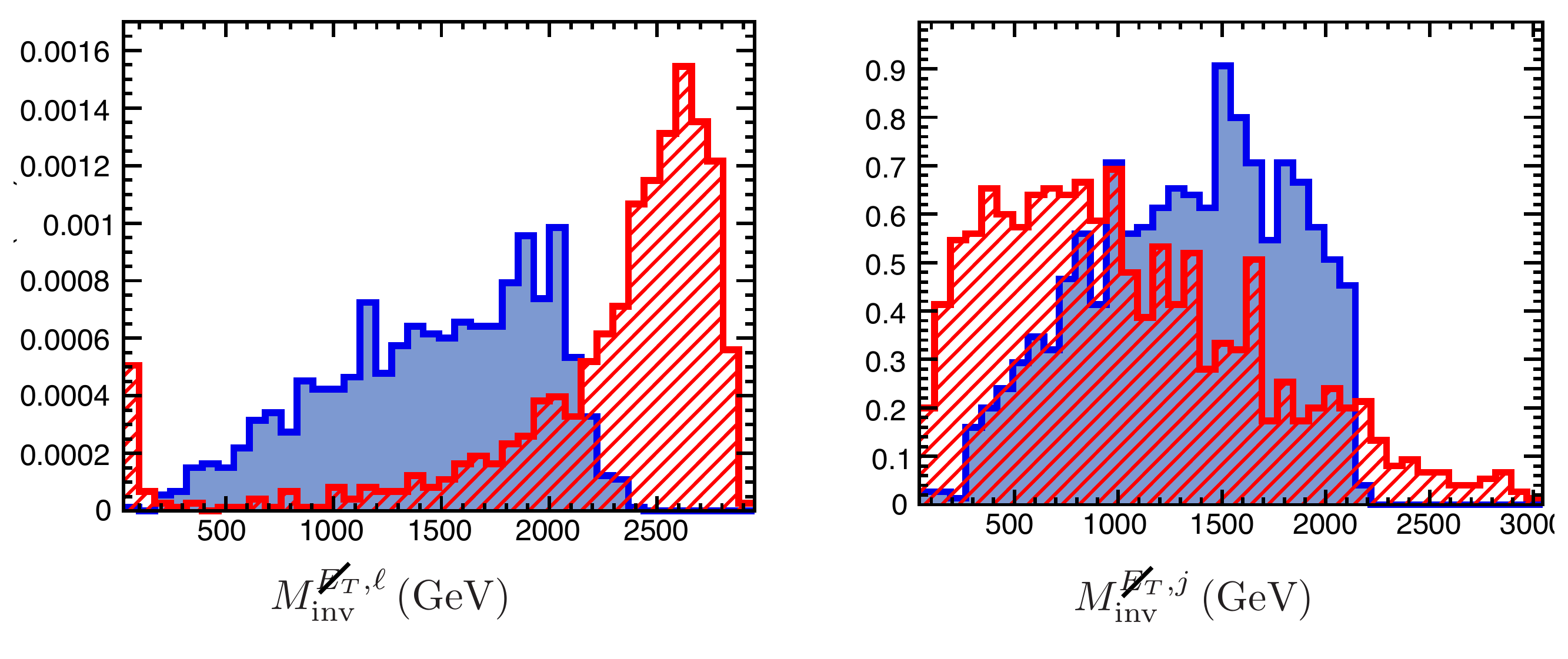}
\caption{\small{ The invariant mass distribution of charged lepton and $\cancel{E_{T}}$  (left panel) and invariant mass distribution between jet and $\cancel{E_{T}}$ (right panel)   for  heavy neutrino mass $M_N=2100$ GeV.}}
\label{2100_fig2}
\end{figure}

Heavy neutrino in the multi TeV mass range can be probed with higher c.m.energy. As an example, we consider $\sqrt{s}=3$ TeV, relevant for CLIC, and present our analysis for the mass range $M_N=1300-2700$  GeV. Similar to the previous analysis, here  we use slightly different cuts for $M_N=1300-1900$ GeV, and $2100-2700$ GeV.  The same set of cuts can not be used for the entire mass range, as the   kinematic  of the final states  for $2700$ GeV are widely different as 1300 GeV. There are few variables that we have taken common though  for both of the regions. These  are electron $p_{T}$, the difference of pseudo-rapidity between jet and MET  $\Delta \eta(j,\cancel{E}_T)$, invariant mass of lepton and MET $M(\cancel{E}_T,\ell)$ and the invariant mass of jet and MET $M(\cancel{E}_T,j)$. We show the distributions of various kinematic variables in Fig.~\ref{2100_fig1} and Fig.~\ref{2100_fig2}. 

\paragraph{} For the mass range 2100-2700 GeV, the electron $e^{\pm}$ from $N$ decay will have very high momentum. Therefore, with stringent cuts on the lepton momentum, the background becomes negligible. We show the distribution for the $p_T$ of lepton in Fig.~\ref{2100_fig1} for the heavy neutrino mass $M_N=2.1$ TeV. We choose a lower $p_T$ cut on electron $p_{T}$  for $M_N=1300-1900$ GeV and larger for the higher mass case. The reason is similar as mentioned for $1.4$ TeV analysis in Section.~\ref{medrange}.  

\paragraph{} In  the right panel of Fig.~\ref{2100_fig1}, we show the distribution of pseudo-rapidity separation between fat-jet and MET.   The separation  is large for large  angular separation.  For relatively lighter  $N$, this  is more likely that the produced fat-jet and $\cancel{E}_T$ have well angular separation between them. Therefore,  we implement a  large cut on $\Delta \eta(j,\cancel{E}_T)$ for $1300-1900$ GeV mass range compared to the  $2100-2700$ GeV  range. For $2100$ GeV mass the peak occurs  around $\Delta \eta(j,\cancel{E}_T)=3.0$. In the  background,  $W^{+}W^{-}$ sample results in a peak around $\Delta \eta(j,\cancel{E}_T)=3.0$.  However,  the background also has other contributions,  that result in  smaller  separation $\Delta \eta(j,\cancel{E}_T)$. Overall the background is more likely to have less angular separation as  compared to the  signal.
The invariant mass distributions for  $3$ TeV, such as,  $M(\cancel{E}_T,\ell)$ and $M(\cancel{E}_T,j)$ have similar  features as  for  $1.4$ TeV. Therefore,  we implement a  strong  cut on these variables for 
relatively lighter  $N$ mass.
Also, $ \Delta \phi(j,\cancel{E}_T)$  is almost   uniformly distributed   for the background, whereas   signal  has  larger cross-section in small $ \Delta \phi(j,\cancel{E}_T)$ region.  Therefore, to enhance the signal sensitivity,  we reject events with  $ \Delta \phi(j,\cancel{E}_T) >  2.0 $.
\begin{figure}[h]
\centering
\includegraphics[width=\textwidth]{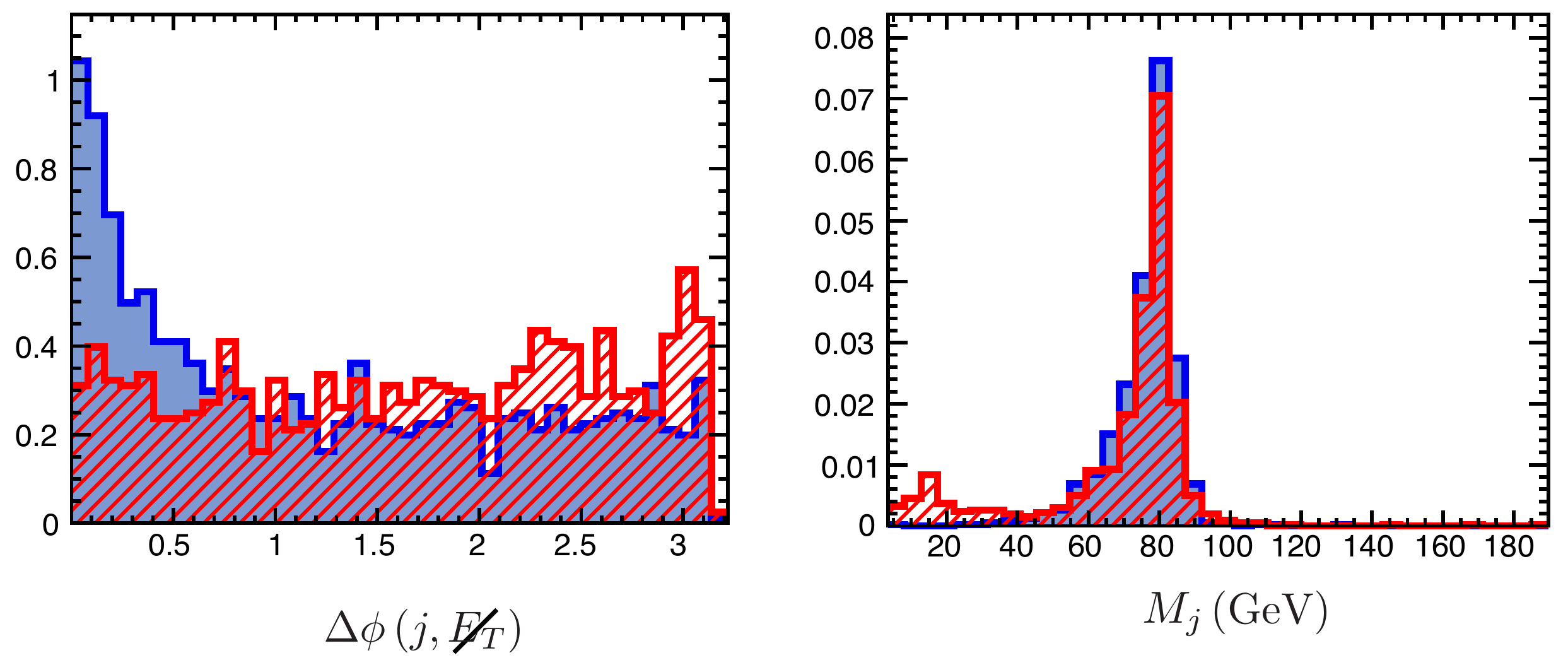}
\caption{\small{ The azimuthal separation between jet and $\cancel{E_{T}}$ (left panel) and distribution of jet mass (right panel) for  heavy neutrino mass $M_N=2100$ GeV. }}
\label{2100_fig3}
\end{figure}
{Additional variable, that we particularly use for 2100-2700 GeV mass range is the jet-mass. }For the signal, jet mass has  a peak near $W$ boson mass as the signal jets are  coming from boosted $W$ boson. Background also has similar peak  around $W$ boson mass, since   $W^{+}W^{-}$ pair production  contributes significantly  in background.  However,  the $W$ boson in the background is relatively less   boosted as compared to  the signal, as this is not generated from the decay of a heavy resonance. This results in a broad peak for the background  compared to the signal. We choose a window  on jet mass variable as $70-90$ GeV. Below, we list all the  cuts that we implement.  Similar to the previous case, the final state contains one isolated lepton $e^{\pm}$, one fat-jet $j_{\rm{fat}}$ with radius $R=1.0$, and missing energy $\cancel{E}_T$. \\

\underline{\textbf{CBA-III for $M_{N}=1300-1900$ GeV}} 
\begin{itemize}
\item $\bf{D1:}$ $p_{T} $ for electron $p_T> 450$ GeV. 
\item $\bf{D2:}$ Pseudo-rapidity of $e^{\pm}$: $-1.0 < {\eta^\ell} < 1.0$.
\item $\bf{D3:}$ Jet-missing energy rapidity separation $\Delta \eta(j,\cancel{E_T})$: $\Delta \eta(j,\cancel{E_T}) > 3.0 $. 
\item $\bf{D4:}$ Jet-lepton rapidity separation $\Delta \eta(j,\ell)$: $ \Delta \eta(j,\ell) < 2.0$. 
\item $\bf{D5:}$ Invariant mass of transverse missing energy and lepton:  $200\, {\textrm{GeV}} <M(\cancel{E}_T,\ell)  < 2500$ GeV. 
\item $\bf{D6:}$ Invariant mass of transverse missing energy and jet:  $M(\cancel{E}_T,j)> 1300.0$ GeV.
\end{itemize}

\underline {\textbf{CBA-IV  for $M_{N}=2100-2700$ GeV}} 
\begin{itemize}
\item $\bf{D1:}$ $p_{T} $ for electron: $p_{T} > 600$ GeV.
\item $\bf{D2:}$ Missing transverse energy: $\cancel{E}_T < 200.0$ GeV.
\item $\bf{D3:}$ Jet-missing energy rapidity separation $\Delta \eta(j,\cancel{E_T})$: $ \Delta \eta(j,\cancel{E_T}) > 0.5 $.
\item $\bf{D4:}$ Jet-missing energy azimuthal angle separation $\Delta \phi(j,\cancel{E_T})$: $ \Delta \phi(j,\cancel{E_T}) < 2.0 $.
\item $\bf{D5:}$ Invariant mass of transverse missing energy and lepton:  $200\, { \rm{GeV}}<M(\cancel{E}_T,\ell) < 2000$ GeV.
\item $\bf{D6:}$ Invariant mass of transverse missing energy and jet: $200.0\,  {\rm{GeV}} < M(\cancel{E}_T,j) < 2000.0$ GeV.
\item $\bf{D7:}$ Jet mass $M_{J}$: $80.0 < M_{J} < 90.0$.
\end{itemize}

Before going into the details of signal and background efficiencies with the full cut based analysis, we discuss the important issues pertaining to MVA and also present a comparative study between the two methods.
After a detailed discussion about the Multivariate analysis, we will  discuss  the results. We also project out the required luminosity to obtain a discovery significance.

\subsection{Multivariate Analysis \label{mva}}
We optimize our search strategy and show the importance of our chosen variables by performing a multivariate analysis using the Boosted Decision Tree (BDT) algorithm. This is implemented within the $\texttt{ROOT}$ framework as Toolkit for Multivariate Analysis (TMVA). In order to classify a set of data, a binary structured decision tree takes yes/no decision on one single variable at a time until some stop criterion is satisfied. Obviously, the classification is whether the data is signal or background like. For example, in our case, the tree starts with a root node and uses variables such as $p_T^{\ell}$, $p_T^j$, $M_{\text{int}}^{\cancel{E_T},j}$, $M_{\text{int}}^{\cancel{E_T},\ell}$, $\cancel{E_T}$, $\eta^{\ell}$ and so on to segregate the data into signal like or background like. A variety of separation criterion can be used to discriminate between the signal and background events. Perhaps, the most common is the Gini index defined by $p\left(1-p\right)$, where $p$ is the purity of the sample. This iteration stops when the maximum separation between signal and background samples are achieved. Extending this concept from one tree to several trees, which eventually forms a forest (random forest), is called boosting. This is extremely important as the outcome of a single decision tree is susceptible to statistical fluctuations. Boosting helps to reduce such errors by giving a larger weight to the misclassified events for the next iteration. Ultimately, the majority vote among the trees in the random forest are taken to classify the events.

\begin{center}
\begin{figure}[h!]
\center
\includegraphics[width=15.3cm, height=7.5cm]{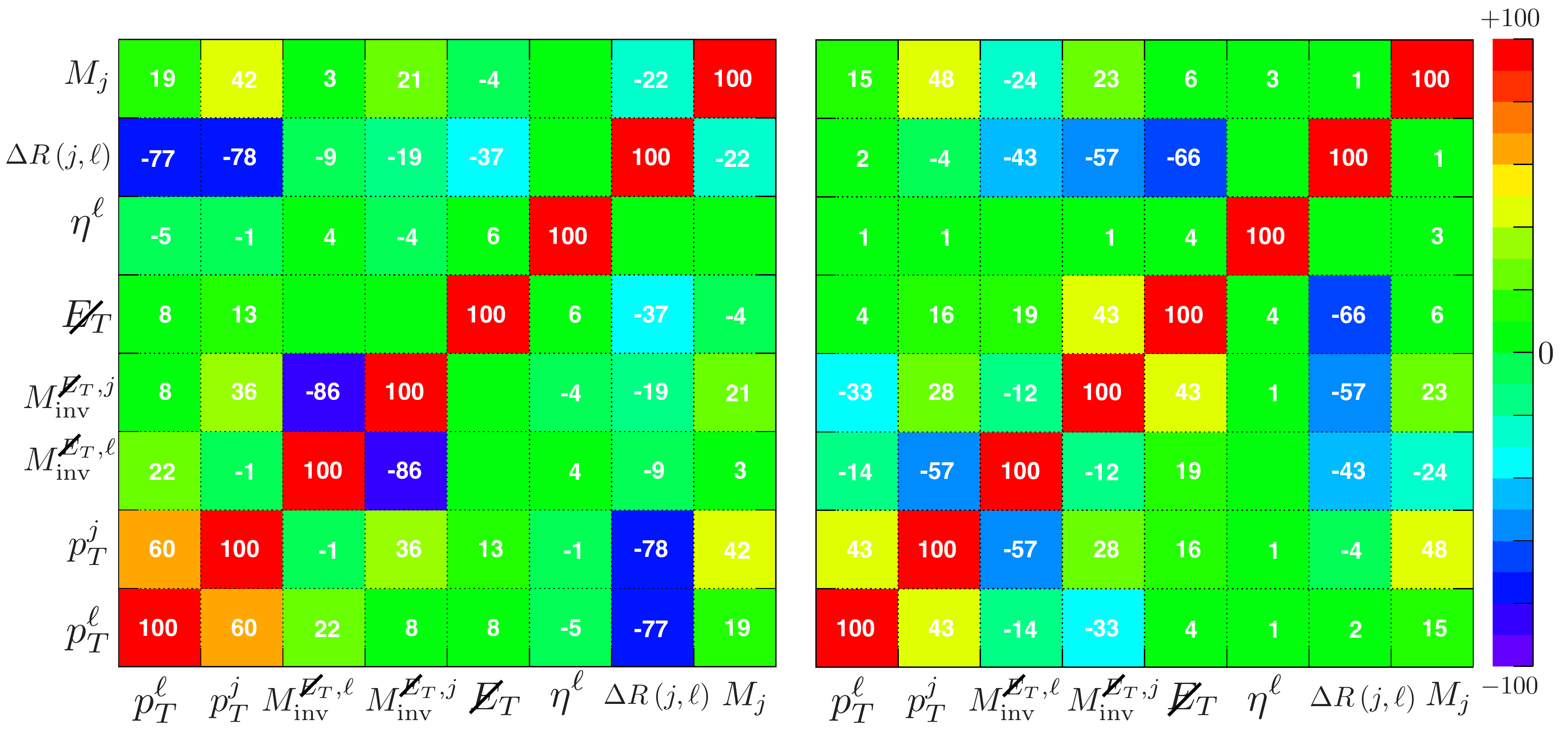}
\caption{{The  plot in the left panel depicts correlation for the used variables for signal events while the plot in the right panel is for background events. We consider $\sqrt{s}=1.4$ TeV.}}
\label{fig:corr_plot}
\end{figure}
\end{center}

For our work, we choose the BDT parameters as: $\texttt{NTrees}$ or the number of trees in a forest to be 400. The maximum depth of the decision tree is considered to be $\texttt{MaxDepth=5}$ and the minimum percentage of training events required in a leaf node is taken as $\texttt{MinNodeSize=2.5\%}$. For boosting the decision tree, we consider the $\texttt{AdaBoost}$ method and the corresponding learning rate for $\texttt{AdaBoost}$ algorithm is taken to be $\texttt{AdaBoostBeta=0.5}$. We also present correlation plots as well as BDT responses using TMVA in Fig.~\ref{fig:corr_plot} and Fig.~\ref{fig:distri3} respectively. The correlation between any two random variables used in our analysis (say $X$ and $Y$) is defined as
\begin{eqnarray}
\rho\left(X,Y\right) &=& \frac{\text{cov}\left(X,Y\right)}{\sigma_X\sigma_Y},
\end{eqnarray}
where $\sigma$ is the usual standard deviation of the input variables and $\text{cov}\left(X,Y\right)\equiv E\left(XY\right)-E\left(X\right)E\left(Y\right)$. It is rather conspicuous that $\rho=0$ would imply independent variables. Usually, the more independent variables are, the more information it carries and therefore helps to distinguish between signal and background events. To quantify the performance of each variable, the relative ranking among the variables are given as: i.) $M_{\text{inv}}^{\cancel{E_T},\ell}$, ii.) $M_{\text{inv}}^{\cancel{E_T}, j}$, iii.) $p_T^{\ell}$, iv.) $p_T^j$, v.) $\cancel{E_T}$, vi.) $\eta^\ell$, vii.) $\eta^j$ and finally viii.) $\Delta R\left(j,\ell\right)$. These ranking or performance of the chosen variables may not be always obvious from the distribution plots shown in Fig.~\ref{fig:distri1}--Fig.~\ref{fig:distrib4}. Hence, ranking of the input variables are obtained based on how often these variables are used to split the decision trees.  The BDT output describes a mapping between the n-dimensional phase space of our chosen variables to a one-dimensional variables. In general, any specific value of the BDT variable can be chosen as a cut, however, a particular cut value in the BDT output corresponds to maximum signal purity and consequently, maximum signal significance. We have also compared our results with the commonly used cut based analysis with the state-of-the-art multivariate analysis. Obviously, significant enhancement in both signal purity and signal significance can be achieved by using MVA.

\subsection{Signal and background efficiency: \label{sensi}}
We divide the discussion of this section into two categories. Firstly, the signal and background significance for $\sqrt{s}={1.4}$ TeV is discussed,  followed by the discussion  for  3 TeV c.m.energy. We also compare our results from  both the cut based and multivariate analysis.
\subsubsection{Signal and background efficiency for $\sqrt{s}=1.4$~TeV:}
As a benchmark, we show the gradual change in the cross-section in Table~\ref{tab:14tevcut1} after implementing the cuts as discussed earlier.

\begin{center}
\begin{table}[h!]
\centering
\begin{tabular}{||c|c|c|c|c|c|c|c|c||}
\hline 
\hline
\multicolumn{1}{|c||}{
Mass (GeV)} & \multicolumn{7}{|c||}{ Cross-sections at the partonic level and  after cuts } \\ \hline \hline 
 \multirow{2}{*}{900} & $\sigma_{partonic}$ (fb) &  C1+C2 & C3 & C4 & C5 & C6 & C7 \\\cline{2-8}

  & 1.78 & 1.24 & 1.01 & 0.88 & 0.85 & 0.83 & 0.73   \\
\hline
\hline
Background & 751.42 & 78.02 & 28.83 & 13.70 & 13.50 & 5.96 & 1.86\\
\hline \hline
\end{tabular} 
\caption{Partonic cross-section  and the cross-section   after each of the cuts for illustrative signal mass point $M_N=900$ GeV.  We also show the background cross-section.} 
\label{tab:14tevcut1}
\end{table}
\end{center}

In Table~\ref{tab:14tevnsiga}, and Table.~\ref{tab:14tevnsigb}, the 2nd column corresponds  to the partonic cross-section ($\sigma_{partonic}$) for  $e^{\pm}+jj+\cancel{E}_T$.   The 3rd column represents the cross-section after all the cuts, where we also include detector effect.   For the mass range $600-1200$ GeV, the partonic cross-section varies in between $\sigma_{partonic} \sim 2.39-0.8$ fb. After taking into account all the above mentioned cuts, and detector effect, the cross-section drops nominally by a factor of $\sigma_D/\sigma_{partonic} \sim 2-3$. For comparatively lower masses, such as, 600 GeV, the drop is relatively larger. This  happens, as  for relatively lower $M_N$,  the decay products  $W^{\pm}$ and charge lepton $l^{\mp}$  have smaller $p_{T}$ as  compared to the higher $M_N$ scenario.  A  high $p_{T}$ $W^{\pm}$-boson have larger probability to make fatjet compared to the low $p_{T}$ $W^{\pm}$-boson. Therefore, for higher $M_N$,  the cuts reduce the signal cross-section nominally.  On the other hand, the background cross-section  $\sigma_{BKG} \sim 751 $ fb at the partonic level, falls drastically $\sigma_{BKG} \sim 1.86$ fb after all the cuts. In particular, we stress that the choice of a high $p_T$ for lepton and jet kills almost all of the backgrounds. 
\begin{table}[h!]
\centering
\begin{tabular}{||c|c|c|c|c||}
\hline 
\multicolumn{3}{|c||}{ Mass and cross-section } & \multicolumn{2}{|c||}{Significance} \\ \hline \hline
Mass (GeV) & $\sigma_{partonic}$ (fb) & $\sigma_{D}$ (fb) & CBA-I  & BDT  \\
\hline
600 & 2.39 & 0.63 & 8.92 & 13.05  \\
700 & 2.24 & 0.77 & 10.61  & 14.06 \\
800 & 2.03 & 0.82 &  11.20  & 14.15  \\
900 & 1.78 & 0.73 &  10.14 & 13.22 \\
\hline
Background & 751.42 & 1.86 & - & -\\
\hline \hline
\end{tabular} 
\caption{Cross-section for signal and background in fb.  We also show the  significance  for luminosity $500\,  \rm{fb}^{-1}$.} 
\label{tab:14tevnsiga}
\end{table}

{The signal sensitivity can be computed  using  the following  expression:
\begin{equation}
n_s=\frac{S_d}{\sqrt{S_d+B_d}}
\label{eq:significance}
\end{equation}
where $S_d$ and $B_d$ represent the signal and background event numbers after all the cuts and detector effect.}
We show the signal sensitivity  in 4th and 5th column of Table.~\ref{tab:14tevnsiga}, and Table.~\ref{tab:14tevnsigb}.   For both the lower and higher masses,  the  significance is  lower and peaks  in the middle region.  For  lower mass of $N$, the cross section is larger and for  higher mass cross-section  is smaller. However,  the cut efficiency is low for small masses, that result in the drop of signal cross-section. The fall of cross-section and sensitivity in higher mass regime occurs due to smaller partonic 
cross-section.  Significance curve for BDT and cut based have similar  features.

\begin{table}[h!]
\centering
\begin{tabular}{||c|c|c|c|c||}
\hline 
\multicolumn{3}{|c||}{ Mass and cross-section } & \multicolumn{2}{|c||}{Significance} \\ \hline \hline
Mass (GeV) & $\sigma_{partonic}$ (fb) & $\sigma_{D}$ (fb) & CBA-II  & BDT  \\
\hline
1000 & 1.49 & 0.62 & 9.41 & 12.49  \\
1100 &  1.16 & 0.51 & 7.94  & 11.41 \\
1200 & 0.80 & 0.30 & 4.93  & 8.61  \\
\hline
Background & 751.42 & 1.55 & - & - \\
\hline \hline
\end{tabular} 
\caption{Cross-section for signal and background in fb.  We also show the  significance  for luminosity $500\,  \rm{fb}^{-1}$. }
\label{tab:14tevnsigb}
\end{table}

\begin{figure}[h!]
\centering
\includegraphics[width=\textwidth,height=7cm]{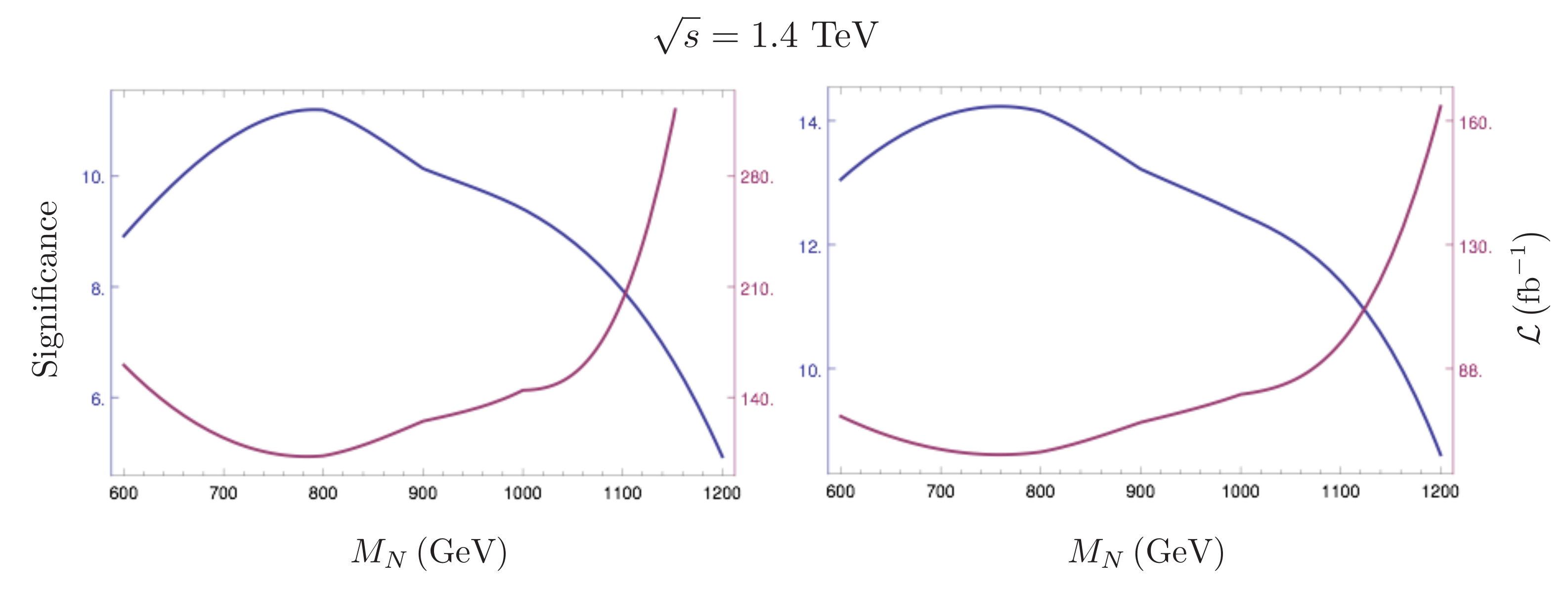}
\caption{\small{{Left:  Variation of signal significance  vs mass of the heavy neutrino using cut-based analysis at $\sqrt{s}=\, 1.4\, \rm TeV$. Right: Same plot as left using BDT. We also show the  required luminosity to achieve $5\sigma$ significance.  The active-sterile mixing has been considered $|V_{eN}|=0.01$.}}}
\label{fig:distri3}
\end{figure}

\subsubsection{Signal and background efficiency for $\sqrt{s}=3$~TeV:}
We discuss the results obtained using both the cut based analysis and MVA   for $\sqrt{s}=3$~TeV.

\begin{center}
\begin{table}[h!]
\centering
\begin{tabular}{||c|c|c|c|c|c|c|c|c||}
\hline 
\multicolumn{1}{|c||}{
Mass (GeV)} & \multicolumn{8}{|c||}{ Cross-sections at the partonic level and  after cuts } \\ \hline \hline 
 \multirow{2}{*}{2100} & $\sigma_{partonic}$ (fb) &  D1 &  D2 & D3 & D4 & D5 & D6 & D7 \\\cline{2-9}
 & 1.61 & 0.91 & 0.76 & 0.76 & 0.54 & 0.49 & 0.47 & 0.38   \\
\hline
\hline
Background & 472.5 & 27.29 & 7.02 & 5.95 & 3.57 & 1.89 & 1.70 & 1.38\\
\hline \hline
\end{tabular} 
\caption{Partonic cross-section  and the cross-section   after each of the cuts for illustrative signal mass point $M_N=2100$ GeV.  The cross-section for background has also been shown.}
\label{tab:14tevcut}
\end{table}
\end{center}

The cross-section for the signal and the background is given in Table.~\ref{tab:3tevnsiga}, and in Table.~\ref{tab:3tevnsigb}, for the mass ranges 1300-1900 GeV, and 2100-2700 GeV, respectively. Similar to the previous 
analysis, the 2nd, and 3rd  column represent the partonic cross-section, and cross-section after all the cuts. For the above mentioned mass range, the partonic cross-section varies in between $\sigma_{partonic} \sim 2.48-0.60$ fb.  
The background cross-section for $\sqrt{s}=3$ TeV c.m.energy is  $\sigma_{BKG}=472.36 $ fb at the partonic level,  and drops down to  sub-fb level after all the cuts. For the signal, the effect of the cuts are nominal, reducing 
the cross-section to $\sigma_D=0.78-0.17 $ fb. A detailed  cut-efficiency is presented in   Table.~\ref{tab:14tevcut},  for the illustrative  signal sample $M_N=2100$ GeV and also for the background.

\begin{table}
\centering
\begin{tabular}{||c|c|c|c|c||}
\hline 
\multicolumn{3}{|c||}{ Mass and cross-section } & \multicolumn{2}{|c||}{Significance} \\ \hline \hline
Mass (GeV) & $\sigma_{partonic}$ (fb) & $\sigma_{D}$ (fb)  & CBA-III  & BDT  \\
\hline
1300  & 2.48 & 0.78 & 13.41 & 16.15  \\
1500 & 2.33 & 0.81 & 13.81  & 18.61 \\
1700 & 2.12 & 0.71 & 12.47  & 19.60  \\
1900 & 1.89 & 0.55 & 10.17 & 17.89 \\
\hline
Background & 472.5 & 0.91 & - & - \\
\hline \hline
\end{tabular} 
\caption{Cross-section for signal and background in fb.  We show the significance for  luminosity $500\,  \rm{fb}^{-1}$. }
\label{tab:3tevnsiga}
\end{table}

\begin{table}
\centering
\begin{tabular}{||c|c|c|c|c||}
\hline 
\multicolumn{3}{|c||}{ Mass and cross-section } & \multicolumn{2}{|c||}{Significance} \\ \hline \hline
Mass (GeV) & $\sigma_{partonic}$ (fb) & $\sigma_{D}$ (fb)  & CBA-IV  & BDT  \\
\hline
2100 & 1.61 & 0.38 & 6.40 & 16.53  \\
2300 & 1.31 & 0.36 & 6.10  & 16.46 \\
2500 & 0.97 & 0.27 & 4.70  & 14.99  \\
2700 & 0.60 & 0.17 & 3.05 & 10.86 \\
\hline
Background & 472.5 & 1.38 & - & - \\
\hline \hline
\end{tabular} 
\caption{Cross-section for signal and background in fb.  We show the significance for  luminosity $500\,  \rm{fb}^{-1}$. } 
\label{tab:3tevnsigb}
\end{table}

\begin{figure}[h!]
\centering
\includegraphics[width=\textwidth,height=7cm]{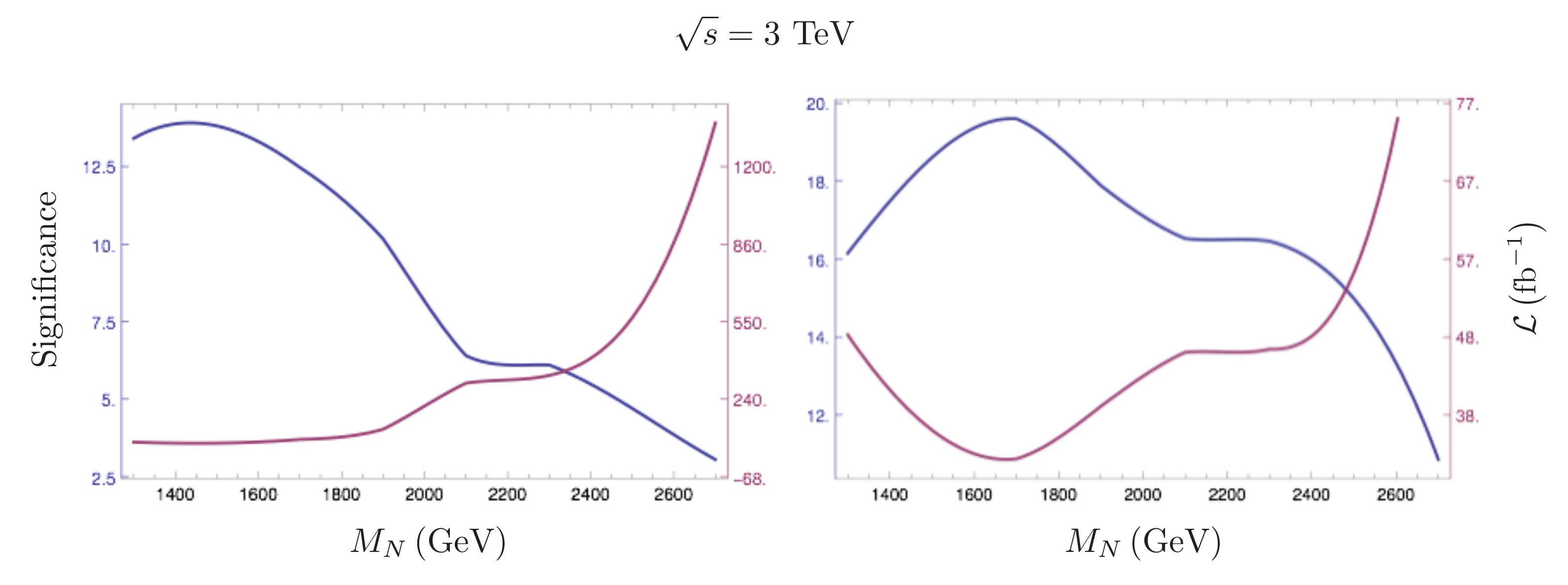}
\caption{\small{{Left:  Signal significance  vs mass of the heavy neutrino using cut-based analysis at $E_{cm}=\, 3 \rm TeV$. Right: Same plot as left using BDT. We also show the  required luminosity to achieve $5\sigma$ significance.  The active-sterile mixing has been considered $|V_{eN}|=0.01$.}}}
\label{fig:distri4}
\end{figure}

In Fig.~\ref{fig:distri3} and Fig.~\ref{fig:distri4},  we  show the variation of  signal sensitivity with mass $M_N$. Both the figures have similar feature. For lower value of $M_{N}$,  the cut efficiency is low, that results in smaller signal cross-section and reduced sensitivity. For higher mass, the reduction occurs due to lower partonic cross-section. The signal  significance reaches maximum  in the mid region. {We also show the required luminosity to achieve $5\sigma$ significance in the same plot.} We emphasise that, heavy neutrino in the  mass range $M_N= 600-1100$ GeV can be discovered  with  $\mathcal{L} \le 100 \,   \rm{fb}^{-1}$ of data in the $\sqrt{s}=1.4 $ TeV run of CLIC. For the c.m.energy 3 TeV,  the required luminosity to probe $M_N= 1300-2300 $ GeV is $\mathcal{L}= 50 \,  \rm{fb}^{-1}$ .   
\begin{center}
\begin{figure}[h!]
\centering
\includegraphics[width=0.75\textwidth]{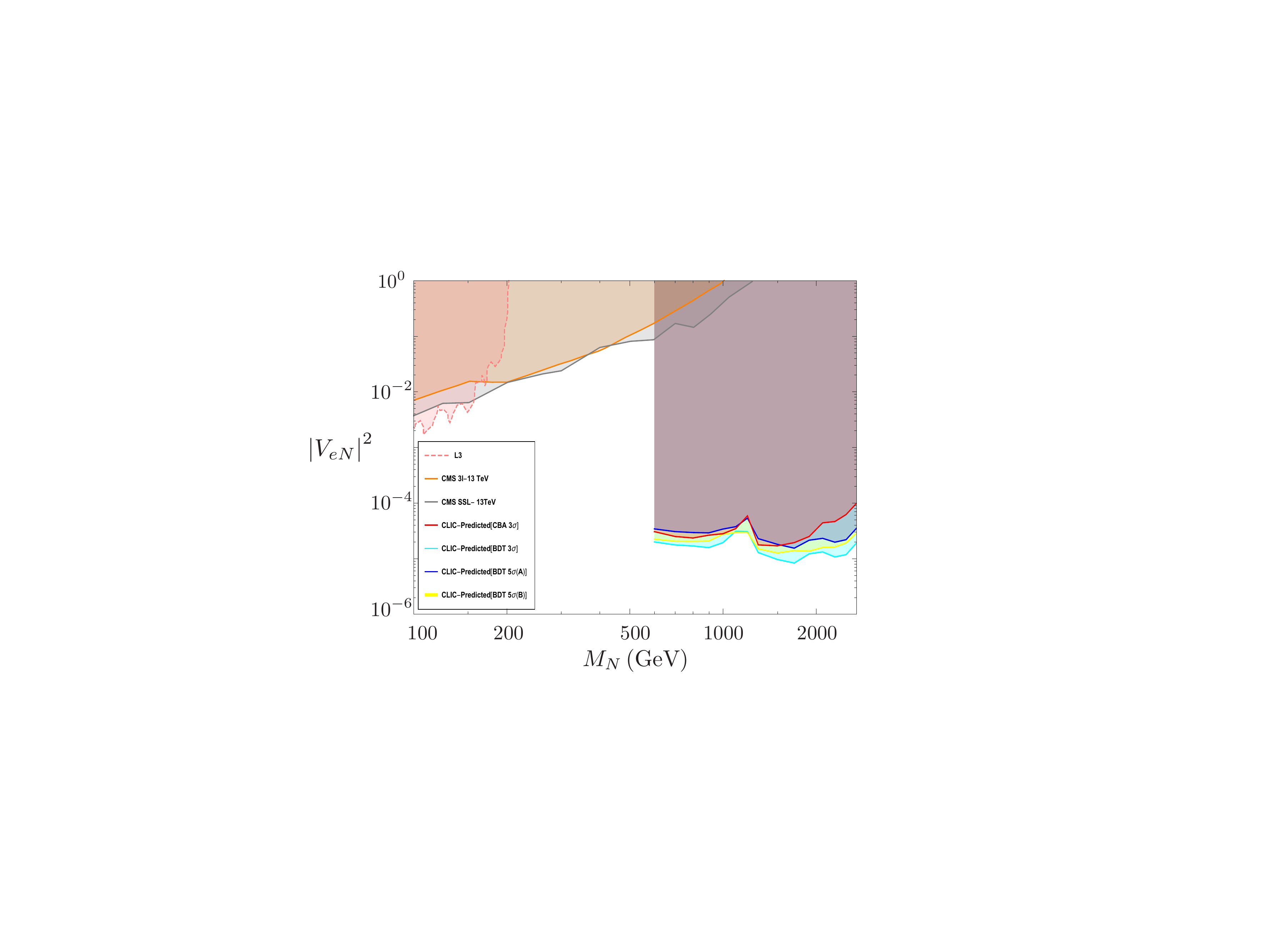}
\caption{\small{Limit on the active-sterile  mixing vs mass of the heavy neutrino. The different bounds correspond to  the CMS $3\ell+\cancel{E}_T$ search \cite{Sirunyan:2018mtv}, CMS $2\ell+jj$ \cite{Sirunyan:2018xiv}, the limit from LEP \cite{Achard:2001qv}. {The limits from cut-based and BDT are in agreement with Table.~\ref{tab:14tevnsiga}, Table.~\ref{tab:14tevnsigb}, Table.~\ref{tab:3tevnsiga}, and Table.~\ref{tab:3tevnsigb}. CLIC-Predicted [CBA $3\sigma$] and CLIC-Predicted [BDT $3\sigma$] lines represents the $3\sigma$ limit, obtained using cut based and BDT analysis respectively. These two limits have been derived without using jet-lepton invariant mass variable as an input of BDT. CLIC-Predicted [BDT $5\sigma$(A)] represents the $5\sigma$ sensitivity and it has also been derived without jet-lepton invariant mass variable. CLIC-Predicted [BDT $5\sigma$(B)] corresponds to $5\sigma$ sensitivity, where in addition to other variables, jet-lepton invariant mass has also been used.} }}
\label{fig:mass_mix_new}
\end{figure}
\end{center}
{For the previous discussions, we considered a benchmark value for the active-sterile mixing $|V_{eN}|=0.01$. The cross-section for the heavy neutrino production varies quadratically with the mixing. Hence, using Eq.~(\ref{eq:significance}), the bound on the active-sterile mixing can be obtained as follows: 
\begin{equation}
n_{s} = \frac{\sigma^{0}_{s}\rm |V_{eN}|^{2}\sqrt{\mathcal{L}}}{\sqrt{\sigma^{0}_{s} \rm |V_{eN}|^{2}+\sigma_{B}}}.
\end{equation}
In the above,  $\sigma^{0}_{s}$ is the signal cross section for unit mixing, and  $\sigma_{B}$ is the background cross section.   $\mathcal{L}$ is the required luminosity to achieve   $n_{s} \sigma$ significance.  Using the above equation, we derive the bound on active-sterile mixing, that we show in Fig.~\ref{fig:mass_mix_new}. We  consider  $\mathcal{L}= 500\, \rm fb^{-1}$, and $n_s=3$.  Similar to the cut-based analysis, we also show the bounds for BDT analysis. Note that, the bound from BDT is factor of $3$ stronger than the cut-based analysis. We find that a  heavy neutrino of mass $900-1200$ GeV and mixing $|V_{eN}|^2=(2.8-5.3) \times 10^{-5}$ can be discovered with $5\sigma$ significance ($|V_{eN}|^2=(1.5-3.0) \times 10^{-5}$ for $3\sigma$) using $\mathcal{L}=500$ $\rm{fb}^{-1}$ luminosity at $\sqrt{s}=1.4$ TeV c.m.energy. More massive heavy neutrino of mass   $M_N=1700-2700$ GeV and  mixing $|V_{eN}|^2=(1.5-3.5) \times 10^{-5}$ can be discovered with $5\sigma$ significance($|V_{eN}|^2=(0.8-1.1) \times 10^{-5}$ for $3\sigma$) at  $\sqrt{s}=3$ TeV c.m.energy  using  $\mathcal{L}=500$ $\rm{fb}^{-1}$ of data. So far in our analysis we have not used jet-lepton invariant mass cut.  This mass cut enhances the signal significance. As a result this improves the mass vs mixing bound by $5-30\%$. This has been shown in the Fig.~\ref{fig:mass_mix_new} .
For comparison, we also show the present LHC limits.  As can be seen,  the leptonic collider is much more effective than  the hadronic collider to constraint the mixing angle for higher masses. In \cite{Bhardwaj:2018lma}, the authors analysed the discovery prospect of heavy neutrino at HL-LHC, using using sub-structure analysis. For higher masses, the sensitivity reach is $|V_{lN}|^2 \sim 10^{-1}-10^{-2}$. We find that for heavier $N$, the $e^{+}e^{-}$ collider can probe upto much lower value of active-sterile mixing and hence will have better sensitivity reach. 
}

\section{Conclusion \label{conclusion}}
We  explore the discovery prospect of a heavy neutrino with  intermediate and  large mass ranges   $M_N=600-1200 $ GeV, and $1300-2700$ GeV  at  the proposed $e^+e^-$ collider for two different c.m.energies $\sqrt{s}=1.4$ TeV and 3 TeV,  respectively.   The heavy neutrino can be produced at the $e^+e^-$ collider through the $s$ and $t$ channel processes, $e^+ e^- \to \nu_e N$, and decays subsequently. We consider the decay mode with highest branching ratio $N \to e^{\pm} W^{\mp}  $. The produced $W^{\pm}$ gauge bosons are   highly boosted, and hence their decays produce collimated decay products.  We consider the hadronic final states of the produced $W^{\pm}$s, that lead to fat-jet.   The model signature is therefore $e^{\pm} + {j}_{\rm{fat}}+ \cancel{E}_T$.    For the background, we generate the events as $e^{\pm} \nu_e/\bar{\nu}_e j j$, that can come from $W^{\pm } W^{\mp}$ sample, but has also other contributions. 
  
For the $\sqrt{s}=1.4$ TeV  analysis, we use optimised cuts to probe  the mass regions $M_N=600-900$ GeV, and $1000-1200$ GeV.  The charged lepton produced from $N$ has relatively larger $p_T$ for  1000-1200 GeV mass range. The cuts on $p_T$ of leptons, as well as other variables, such as, $\eta^\ell, \Delta R(j,\ell), M(\cancel{E}_T,\ell)$ remove majority of the SM background.  We find that the entire mass range $600-1100$ GeV have fairly large signal cross-section $\sigma_D=0.51-0.82$ fb, after taking into account the detector effect. For the background, the  cross-section falls after all the cuts,  from 751 fb as partonic cross-section to $\sigma_D \sim 1 $ fb. In addition to the cut-based analysis, we also pursue multivariate analysis. We find that the heavy neutrino of mass $M_N=600-1200 $ GeV and the active-sterile mixing $|V_{eN}|^2 \sim  10^{-5}$ can be discovered at $5\sigma$ significance with 500 $\rm{fb}^{-1}$ luminosity. 
  
Similar to this analysis, we also pursue the analysis for $\sqrt{s}=3$ TeV c.m.energy, using the same set of tools.  We explore the mass range $1300-2700$ GeV for this case. For this ultra heavy $M_N$, the produced $e^{\pm}$ and  $W^{\pm}$s are even more boosted. The lepton and fat-jets have very high $p_T$. Typically, for $M_N=2100$ GeV, the peak in $p_T$ occurs around 1000 GeV. We use cuts on different kinematic variables, as well as, followed a MVA prescription. We find that heavy neutrino of mass  $M_N=1300-2700 $ GeV with mixing $|V_{eN}|^2 \sim   10^{-5}$, can be discovered at $5\sigma$ significance with 500 $\rm{fb}^{-1}$ luminosity.

  \section*{Acknowledgements}
MM acknowledges the support of the DST-INSPIRE research grant IFA14-PH-99,  and the HPC cluster facility at IOP, Bhubaneswar. The authors acknowledge the hospitality of IISER Bhopal during WHEPP-XV, where this work has been initiated.

\bibliography{inverse.bib}
\bibliographystyle{JHEP}
\end{document}